\documentclass[preprint,aps,prc,tightenlines,nofootinbib,showpacs]{revtex4}

\begin{document}

\input{psfig}
\input{epsf}
\def\Im{\mbox{\sl Im\ }}
\def\pd{\partial}
\def\oln{\overline}
\def\ola{\overleftarrow}
\def\ora{\overrightarrow}
\def\ds{\displaystyle}
\def\bgreek#1{\mbox{\boldmath $#1$ \unboldmath}}
\def\sla#1{\slash \hspace{-2.8mm} #1}
\newcommand{\bra}{\langle}
\newcommand{\ket}{\rangle}
\newcommand{\vep}{\varepsilon}
\newcommand{\met}{{\mbox{\scriptsize met}}}
\newcommand{\lab}{{\mbox{\scriptsize lab}}}
\newcommand{\cm}{{\mbox{\scriptsize cm}}}
\newcommand{\mcal}{\mathcal}
\newcommand{\Del}{$\Delta$}
\newcommand{\g}{{\rm g}}
\long\def\Omit#1{}
\long\def\omit#1{\small #1}
\def\beq{\begin{equation}}
\def\eeq{\end{equation} }
\def\bea{\begin{eqnarray}}
\def\eea{\end{eqnarray}}
\def\eqref#1{Eq.~(\ref{eq:#1})}
\def\eqlab#1{\label{eq:#1}}
\def\figref#1{Fig.~\ref{fig:#1}}
\def\figlab#1{\label{fig:#1}}
\def\tabref#1{Table \ref{tab:#1}}
\def\tablab#1{\label{tab:#1}}
\def\secref#1{Section~\ref{sec:#1}}
\def\seclab#1{\label{sec:#1}}
\def\VYP#1#2#3{{\bf #1}, #3 (#2)}  
\def\NP#1#2#3{Nucl.~Phys.~\VYP{#1}{#2}{#3}}
\def\NPA#1#2#3{Nucl.~Phys.~A~\VYP{#1}{#2}{#3}}
\def\NPB#1#2#3{Nucl.~Phys.~B~\VYP{#1}{#2}{#3}}
\def\PL#1#2#3{Phys.~Lett.~\VYP{#1}{#2}{#3}}
\def\PLB#1#2#3{Phys.~Lett.~B~\VYP{#1}{#2}{#3}}
\def\PR#1#2#3{Phys.~Rev.~\VYP{#1}{#2}{#3}}
\def\PRC#1#2#3{Phys.~Rev.~C~\VYP{#1}{#2}{#3}}
\def\PRD#1#2#3{Phys.~Rev.~D~\VYP{#1}{#2}{#3}}
\def\PRL#1#2#3{Phys.~Rev.~Lett.~\VYP{#1}{#2}{#3}}
\def\FBS#1#2#3{Few-Body~Sys.~\VYP{#1}{#2}{#3}}
\def\AP#1#2#3{Ann.~of Phys.~\VYP{#1}{#2}{#3}}
\def\ZP#1#2#3{Z.\ Phys.\  \VYP{#1}{#2}{#3}}
\def\ZPA#1#2#3{Z.\ Phys.\ A\VYP{#1}{#2}{#3}}
\def\half{\mbox{\small{$\frac{1}{2}$}}}
\def\quarter{\mbox{\small{$\frac{1}{4}$}}}
\def\nn{\nonumber}
\newlength{\PicSize}
\newlength{\FormulaWidth}
\newlength{\DiagramWidth}
\newcommand{\vslash}[1]{#1 \hspace{-0.5 em} /}
\def\ser{\marginpar{Mod-Ser}}
\def\kun{\marginpar{Mod-Kun}}
\def\fre{\marginpar{Mod-Fre}}
\def\her{\marginpar{$\Longleftarrow$}}
\def\bel{\marginpar{$\Downarrow$}}
\def\abo{\marginpar{$\Uparrow$}}


\title{In-medium meson properties and field transformations}

\author{S. Kondratyuk}
\author{K. Kubodera}
\author{F. Myhrer}
\affiliation{Nuclear Theory Group, 
Department of Physics and Astronomy,
University of South Carolina,
712 Columbia, SC 29208, USA} 

\date{\today}

\begin{abstract}

Since the existing calculations of the effective meson mass in nuclear medium
involve approximations, it is important
to examine whether they satisfy  the general requirement of the
equivalence theorem that the physical observables should be 
independent of the choice of field variables.
We study here consequences of nucleon field transformations. 
As an illustrative case we consider the in-medium effective pion mass
calculated for the s-wave pion-nucleon interaction
in the linear density approximation. 
We demonstrate that  
it is necessary to include the Born term explicitly in order
that the effective pion mass should obey the equivalence theorem.

\end{abstract}


\maketitle


\section{Introduction} \seclab{intro}

Although the behavior of mesons in nuclear medium,
including the possibility of meson condensation, has been extensively studied 
for several decades~\cite{Kis55,Eri66,Mig71,Kap86}, various aspects of the problem 
continue to draw considerable 
attention~\cite{Yab93,Yab94,Tho95,Lee95,Par02,Pan95}.
A number of calculations of effective 
in-medium meson properties have been done based on effective field theory 
approaches, 
such as chiral perturbation theory ($\chi$PT), 
in order to incorporate the basic properties of low-energy hadron interactions 
in a field-theoretic framework.
One of the issues discussed intensively in the literature
is the off-shell invariance of the in-medium effective meson mass
(or, equivalently, its invariance under transformations of
interpolating field variables).
According to the well-known equivalence theorem~\cite{Haa58,Chi61,Col69,Sal70,Lam73}   
any observables, and hence in-medium observables as well,
should be independent of the choice of field variables.
A question of practical importance is whether this requirement is fulfilled 
in actual calculations that involve approximations such as truncations
of the lagrangian, the linear-density approximation, etc.
(see, e.g., \cite{Yab94,Tho95,Lee95,Par02}).
Detailed studies by Thorsson and Wirzba~\cite{Tho95} 
and Lee {\it et al}.~\cite{Lee95} have demonstrated that the in-medium effective meson 
mass calculated in the linear baryon density approximation
and next-to-leading order in $\chi$PT is independent of the choice of meson field 
variables in conformity with the equivalence theorem.
Beyond the linear-density approximation, one needs to take into account the 
possible influence of multi-baryonic terms in the lagrangian as well as nuclear
correlations. This challenging problem has been addressed by several
authors (see, e.g., Refs.~\cite{Eri66,Lee95,Par02,Pan95}), 
and further studies in this direction are certainly warranted.
 
To shed more light on the issue of the off-shell invariance of 
the in-medium meson properties, in this article we examine consequences of
transformations of the nucleon field. Our study is supplementary to those of
Refs.~\cite{Yab94,Tho95,Lee95,Par02} wherein transformations of 
meson fields were discussed. 
We limit ourselves here to the consideration
of an in-medium effective pion mass
calculated in the standard linear density approximation 
of the optical potential theory~\cite{Kis55,Eri66,Ser86}.
Furthermore, we only deal with the s-wave pion-nucleon interaction. 
Although (possible) meson condensation is very likely to be influenced 
by not only s-wave but also p-wave interactions, we believe it is still 
informative 
to study the s-wave contributions alone. 
While s-wave interactions are more relevant to kaon 
condensation~\cite{Kap86}, 
we consider here the pion for simplicity.
This should not affect the main conclusions of this report 
since they follow from
the transformation of the nucleon -- rather than meson -- field.     
The basic nature of the issues involved is expected to reveal itself even 
in the present limited treatment.
 
The pion-nucleon scattering amplitude is calculated here based on
the relativistic chiral lagrangian~\cite{Gas88} at tree level.
In this formalism the in-medium modification of the pion mass
comprises contributions from contact pion-nucleon interactions 
and from the nucleon exchange (Born) term. 
The contact interaction
includes effects of explicit chiral symmetry breaking
through the nucleon sigma-term and has been extensively 
considered in the past~\cite{Yab94,Tho95}.
In this paper we emphasize that, 
in addition to the contact interaction, 
the Born term must be {\em explicitly} taken into account 
in order to ensure the off-shell invariance 
of the effective pion mass. This is shown by evaluating the effective in-medium
mass in two different representations 
of the lagrangian which are connected by a transformation 
of the nucleon interpolating field. 

\section{The effective lagrangian} \seclab{lagr}

Our arguments are based on the second order 
relativistic chiral lagrangian~\cite{Gas88,Bec99} containing 
the pion field $\varphi_\alpha$ (with isovector index $\alpha=1,2,3$) and the 
nucleon field $\psi$.  
The lagrangian reads
\beq
\mathcal{L} = \mathcal{L}_{\pi \pi}^{(2)} + 
\mathcal{L}_{\pi N}^{(1)} +
\mathcal{L}_{\pi N}^{(2)} \, ,
\eqlab{lagr_general}
\eeq
where the superscripts show the order in $Q$, 
with $Q$ being a generic notation for small four-momenta or the pion mass.
The pion lagrangian $\mathcal{L}_{\pi \pi}^{(2)}$ is 
given in terms of the standard non-linear fields, $U$,  
which we expand in powers of the pion field: 
\beq
U = u^2 =
\exp{\left( i\frac{\overrightarrow{\tau} \!\cdot\! \ora{\varphi}}{f}\right)}
= 1+{i \over f} \ora{\tau} \!\cdot\! \overrightarrow{\varphi} -
{1 \over 2f^2} \overrightarrow{\varphi}^2 +\cdots \; ,
\eqlab{u_big}
\eeq  
\beq
u_\mu = i u^{\dagger} (\partial_\mu U) u^{\dagger} =
-{1 \over f} \overrightarrow{\tau} \!\cdot\!
\partial_\mu\overrightarrow{\varphi} + \cdots 
\eqlab{u_mu}
\eeq 
and 
\beq
\chi_+ = M^2 (U^\dagger + U) = 2M^2-{M^2 \over
  f^2}\overrightarrow{\varphi}^2 + \cdots \; ,
\eqlab{chi_plus}
\eeq
where $f$ is the pion decay constant,
$M$ is the pion mass and $\tau^\alpha$ are the
Pauli isospin matrices.\footnote{We will put arrows over isospin
vectors and use the boldface for the usual three-vectors.}
Since we will consider only tree-level contributions 
to the pion-nucleon scattering amplitude, 
we will omit polynomials containing three or more pion
fields; such terms are denoted by an ellipsis.
The purely pionic lagrangian in \eqref{lagr_general} is
\beq
 \mathcal{L}_{\pi\pi}^{(2)} = {f^2 \over 4}  
\mbox{Tr}\left( u_\mu u^\mu + \chi_+ \right) 
= 
{1 \over 2} (\pd_\mu \ora{\varphi}) \!\cdot\! (\pd^\mu \ora{\varphi}) - 
{M^2 \over 2} \ora{\varphi}^2 +\cdots \; ,
\eqlab{lagr_pion_expl}
\eeq
with the trace pertaining to the $2 \times 2$ isospin matrices. 
The pion-nucleon lagrangian in \eqref{lagr_general} 
involves the covariant derivative    
\beq
D_\mu \psi = \left( \pd_\mu + {1 \over 2} 
[u^\dagger,(\pd_\mu u)] \right) \psi
= \left (\pd_\mu + {i \over 4 f^2} \epsilon_
{\alpha \beta \gamma} \tau^\gamma
\varphi^\alpha (\pd_\mu \varphi^\beta) \right) \psi +\cdots \; ,
\eqlab{covar_deriv}
\eeq 
where we have used \eqref{u_big} for the expansion of $u$. 
The relevant parts of the lagrangian are 
\bea
\mathcal{L}_{\pi N}^{(1)} &=& \oln{\psi} (i \sla{D} - m) \psi + 
{1 \over 2} g_A \oln{\psi} \vslash{u} \gamma_5 \psi 
\nonumber \\ &=& 
\oln{\psi} (i \vslash{\pd} - m) \psi - {1 \over 4f^2} 
\epsilon_{\alpha \beta \gamma} \oln{\psi} \tau^\gamma \varphi^\alpha
(\vslash{\pd} \varphi^\beta) \psi - {g_A \over 2f} \oln{\psi}
\ora{\tau} \!\cdot\! (\vslash{\pd} \ora{\varphi}) \gamma_5 \psi +\cdots \; 
\eqlab{lagr_nucl1_expl}
\eea 
and
\bea
\mathcal{L}_{\pi N}^{(2)} & = & 
c_1 \mbox{Tr}(\chi_+) \oln{\psi}\psi - {c_2 \over 4 m^2} \mbox{Tr}(u_\mu u_\nu) 
\left(\oln{\psi} D^\mu D^\nu \psi + \mbox{herm.~conj.} \right) \nn \\ 
& & + {c_3 \over 2} \mbox{Tr}(u_\mu u^\mu)\oln{\psi}\psi - 
{c_4 \over 4} \oln{\psi} \gamma^\mu \gamma^\nu [u_\mu,u_\nu]\psi \nn \\ 
& = & c_1\left( 4M^2 - {2 M^2 \over f^2} \ora{\varphi}^2 \right)\oln{\psi}
\psi - 
{c_2 \over 2m^2 f^2} (\pd_\mu \ora{\varphi}) \!\cdot\! (\pd_\nu
\ora{\varphi})\oln{\psi}(\pd^\mu \pd^\nu \psi) \nn \\
& & - {c_2 \over 2m^2 f^2} (\pd_\mu \ora{\varphi}) \!\cdot\! (\pd_\nu
\ora{\varphi})(\pd^\mu \pd^\nu \oln{\psi})\psi \! + \!
{c_3 \over f^2}(\pd_\mu \ora{\varphi}) \!\cdot\! (\pd^\mu \ora{\varphi})
\oln{\psi}\psi \! \nn \\ 
& & - \! i \epsilon_{\alpha \beta \gamma} 
{c_4 \over 2 f^2}
\oln{\psi} \tau^\gamma 
(\vslash{\pd}\varphi^\alpha)(\vslash{\pd}\varphi^\beta)\psi +\cdots \; ,   
\eqlab{lagr_nucl2_expl}
\eea
where $m$ is the nucleon mass, $g_A$ is the axial coupling
constant and $c_1, \cdots , c_4$ are the standard 
low energy constants.\footnote{We use the definitions of~\cite{Ber93,Tho95} 
for the low-energy constants, and the conventions of~\cite{Bjo64} 
for the Dirac matrices, Lorentz vectors and their products, Feynman rules, etc.} 

It is convenient to combine $-m \oln{\psi}\psi$ from \eqref{lagr_nucl1_expl}
with $4 c_1 M^2 \oln{\psi}\psi$ from \eqref{lagr_nucl2_expl} to form a
modified nucleon mass $m_N = m - 4 c_1 M^2$. 
The difference between $m_N$ and $m$ will lead to corrections of
$\mathcal{O}(Q^4)$ in all the quantities calculated below.
Since such corrections are beyond the order $\mathcal{O}(Q^2)$ we are interested 
in, we will simply keep the notation $m$ for the nucleon mass and 
drop the term $4 c_1 M^2 \oln{\psi}\psi$ from $\mathcal{L}^{(2)}_{\pi N}$.

\section{Effective pion mass in nuclear matter}\seclab{effmass_rep1}

The effective in-medium pion mass is determined by the pole position of
the full in-medium pion propagator~\cite{Mig71}. 
The pole position corresponds to the energy 
of a pionic mode state and is related to the pion self-energy in the medium. 
We will apply the usual mean-field approximation~\cite{Eri66,Ser86}, 
where, to first order in the nuclear density $\rho$, 
the s-wave pion self-energy is proportional to the 
isospin-symmetric forward pion-nucleon scattering amplitude. 
The general structure of the pion-nucleon scattering 
amplitude is~\cite{Hoh83}  
\beq
T_{\alpha \beta}(p,q;p',q') = \oln{u}(p') \left[ \delta_{\alpha \beta}
\left( D^+ + {[\vslash{q},\vslash{q}'] \over 4m} B^+ \right) 
+{[\tau_\alpha,\tau_\beta]\over 2}                   
\left( D^- + {[\vslash{q},\vslash{q}'] \over 4m} B^- \right) \right] u(p) \, ,
\eqlab{ampl_gen}
\eeq
where the initial and final nucleon and pion
four-momenta are $p$, $p'$ and $q$, $q'$, respectively, 
and the initial and final
pion isospin indices are $\alpha$ and $\beta$. 
The nucleons are on the mass shell -- described by
Dirac spinors $u(p)$, $\oln{u}(p')$ -- while the pions
can be off-shell in general. 
The amplitudes $D^{\pm}$ and $B^{\pm}$ are functions of the 
standard kinematic variables
\beq
\nu=\frac{q^2-q^{\prime 2} + 2(p \, (q+q'))}{4m},\;\; 
\nu_B=-{(q q') \over 2m}.
\eqlab{nu_nub}
\eeq
  
The isospin-symmetric, s-wave interaction is described by the
invariant amplitude $D^+$.  
In the tree-level approximation, the amplitude
comprises contributions from the s- and u-channel nucleon pole 
graphs, which constitute the ``pseudovector'' Born term
$D^+_{Born} = D^+_s + D^+_u$, and from the four-particle 
contact graph, $D_c^+$:
\beq
D^+ = D^+_{Born} + D^+_c = D^+_s + D^+_u + D^+_c \,, 
\eqlab{d_gen}
\eeq  
where the explicit expressions are obtained using the lagrangians  
Eqs.~(\ref{eq:lagr_nucl1_expl}, \ref{eq:lagr_nucl2_expl}):
\bea
D^+_s &=& \frac{m g_A^2}{2 f^2} \left( \frac{\nu_B}{\nu_B - \nu} -
{\nu \over 2m} \right) \,,
\eqlab{d_s}
\\
D^+_u &=& \frac{m g_A^2}{2 f^2} \left( \frac{\nu_B}{\nu_B + \nu} +
{\nu \over 2m} \right) \,,  
\eqlab{d_u}
\\
D^+_c &=& -{4 c_1 M^2 \over f^2} + {c_2 \over m^2 f^2} 
\bigg[ (q p)(q' p) + 
(q p')(q' p') \bigg] + {2 c_3 \over f^2}(q q')\, . 
\eqlab{d_c}
\eea

In the mean-field approximation~\cite{Eri66,Ser86}
the pion self-energy $\Pi(\omega,{\bf k},\rho)$ is given by the 
forward scattering amplitude $D^+$ in the nucleon rest frame, 
i.e., for $\nu$ and $\nu_B$ evaluated from Eqs.~(\ref{eq:nu_nub}) with
$p=p'=(m,{\bf 0})$ and $ q=q'=(\omega,{\bf k})$: 
\beq
\Pi(\omega,{\bf k},\rho) = -\rho \, D^+(\nu=\omega,\nu_B=({\bf k}^2-
\omega^2)/(2m)) \, . 
\eqlab{pion_se_gen}
\eeq
This kinematical situation describes a pion interacting 
with a heavy, non-recoiling nucleon 
in the Fermi sea.
Substitution of Eqs.~(\ref{eq:d_gen}--\ref{eq:d_c}) 
into \eqref{pion_se_gen} yields
\beq
\Pi(\omega,{\bf k},\rho) = {\rho \over f^2} \left[ \frac{g_A^2
\left({\bf k}^2-\omega^2\right)^2}{4m \omega^2} - \sigma -
2(c_2+c_3)\omega^2 + 2 c_3 {\bf k}^2 \right] + \mathcal{O}(Q^3,\rho^2)\,,  
\eqlab{pion_se_expl}
\eeq  
where $\sigma = -4 c_1 M^2$ equals the nucleon sigma-term up
to corrections of $\mathcal{O}(Q^3)$~\cite{Ber93}.

The effective in-medium pion mass $M_{eff}$ is defined as the
pole of the pion propagator
\beq
\left( \omega^2 - {\bf k}^2 - M^2 -  
\Pi(\omega,{\bf k},\rho) \right)^{-1}\,, 
\eqlab{prop}
\eeq
when ${\bf k}={\bf 0}$, i.e., $ M_{eff}$ is the solution
of the dispersion equation for $ \omega $: 
\beq
\omega^2 - M^2 - \Pi(\omega,{\bf k}={\bf 0},\rho) = 0 \, . 
\eqlab{disp_eq}
\eeq  
Using the pion self-energy of \eqref{pion_se_expl}, we get
for the square of the effective mass 
\bea
M_{eff}^2 &=& M^2 + M_{eff, Born}^2 + M_{eff, c}^2 + \mathcal{O}(Q^3,\rho^2) \nn 
\\
&=& M^2 \left[ 1 - \frac{\rho}{f^2 M^2 } 
\left( -\frac{g_A^2 M^2 }{4m} + \sigma + 2(c_2+c_3)M^2 \right) \right] 
+ \mathcal{O}(Q^3,\rho^2) \, , 
\eqlab{meff}
\eea 
where we have separated the Born contribution for 
later convenience. 
The contribution of the Born term, $D^+_{Born}$, to $M_{eff}^2$ is
\beq
M_{eff, Born}^2 = \rho \, \frac{g_A^2 M^2 }{4mf^2}\,, 
\eqlab{meff_born}
\eeq 
while the contribution of the contact term $D^+_c$ equals
\beq
M_{eff, c}^2=-{\rho \over f^2} \left[ \sigma + 2(c_2+c_3)M^2 \right].
\eqlab{meff_ct}
\eeq 
The Born contribution is estimated to be  
$M_{eff, Born}^2 \approx  0.066 \, M^2 (\rho/\rho_0)$,         
where $\rho_0$ is the normal nuclear matter density
($\rho_0=0.17\, \mbox{fm}^{-3}$), using 
$g_A=1.27$, $f=92.4\, \mbox{MeV}$, $M=139\, \mbox{MeV}$,
$m=938\, \mbox{MeV}$. 
However, a 
quantitative evaluation of 
\eqref{meff} is quite sensitive to
the precise values of the sigma-term $\sigma$ and the low-energy constants
$c_2$ and $c_3$ (see  Refs.~\cite{Ber93,Fet00} regarding 
the values of the low-energy constants, and Refs.~\cite{Sai02} 
for a recent discussion on the sigma-term).
Such numerical analysis is outside the scope 
of this work, and we remark only that 
the Born and the contact term contributions to $M_{eff}^2$ 
are of the same order, $\mathcal{O}(Q^2)$.
 
At first sight, the presence of the Born contribution $M_{eff, Born}^2$ 
in \eqref{meff} may not look significant,
as its effect can formally be absorbed into the contact term by using, e.g.,  
${c}_2' = c_2 - g_A^2/(8m)$ instead of $c_2$. 
In fact, the expression $M_{eff}^2 = M^2 + M_{eff, c}^2\,$  
has appeared in the literature~\cite{Tho95}.
We shall show however that, 
in order to ensure that $M_{eff}^2$ is independent of the nucleon
interpolating field, 
one {\em must} keep track of the terms of different origins.

\section{Transformation of the nucleon field and the equivalence 
theorem for the in-medium pion mass}\seclab{transf}

Insofar as the effective pion mass has the status
of an observable quantity, the equivalence theorem requires it
to be invariant with respect to redefinitions of the interpolating -- or off-shell --
fields.
The off-shell invariance of the scattering matrix has been studied 
in various approaches, including axiomatic field theory~\cite{Haa58}, 
lagrangian and hamiltonian formalisms~\cite{Chi61,Col69},
the path-integral method~\cite{Sal70} (which in $\chi$PT is exemplified
by the generating functional approach~\cite{Gas84,Gas88}) 
and Feynman-graph techniques~\cite{Lam73}.
The role of the equivalence theorem in hadronic physics has been 
also discussed recently within several models and approximation schemes~\cite{Sch95}.

The main purpose of the present paper is to study consequences of the
equivalence theorem for the effective in-medium pion mass as calculated using the commonly
employed mean-field approximation.
It should be emphasized that the off-shell invariance of observables holds not only in
general, but also at tree level separately~\cite{Col69}.
This fact is crucial for our present calculation 
as it allows us to apply the equivalence theorem to the tree-level amplitudes,
which significantly simplifies the approach.
It is also important that the equivalence theorem is applicable 
to the transformations involving only
the pion or only the nucleon fields, as well as to more complicated
transformations in which both pion and nucleon field variables change.
The invariance of the effective mass with respect to field transformations
of the pion field separately
was extensively discussed before (see, e.g., Refs.~\cite{Yab94,Tho95,Lee95,Par02}).
Here we consider a complementary case where it is the nucleon field that
undergoes a transformation while the pion field does not change.
Specifically, we consider a transformation of the nucleon field
\beq
\psi \longrightarrow \left( 1 + i {\lambda \over f} \gamma_5
\ora{\tau}\!\cdot\!\ora{\varphi} 
- {\lambda^2 \over 2 f^2} \ora{\varphi}^2
\right) \psi,
\eqlab{transf}
\eeq
where $\lambda$ is a real parameter. This may be viewed as a 
truncated form of
the chiral rotation $\psi \rightarrow \exp\{i (\lambda/f) \gamma_5
\ora{\tau}\!\cdot\!\ora{\varphi}\}\psi$ 
in which the terms linear and quadratic in the pion field are retained.
As before, we drop the terms with three or more pion fields since they are 
irrelevant to our tree-level calculation.
The truncation, however, should not suggest that the parameter of
the transformation is small: the equivalence theorem
holds for any $\lambda$ in \eqref{transf}.
The redefinition of the nucleon field \eqref{transf} entails the
transformation of the lagrangian in \eqref{lagr_nucl1_expl}:
\bea
\mathcal{L}_{\pi N}^{(1)} & \longrightarrow & 
\mathcal{L}_{\pi N}^{(1)} - 
{\lambda \over f} \oln{\psi} \ora{\tau} \!\cdot\! (\vslash{\pd} \ora{\varphi}) 
\gamma_5 \psi
-i{2 m \lambda \over f} \oln{\psi} \ora{\tau} \!\cdot\! \ora{\varphi} \gamma_5 
\psi+
{g_A \lambda \over f^2}\epsilon_{\alpha \beta \gamma} 
\oln{\psi}(\vslash{\pd}\varphi^\alpha)
\varphi^\beta \tau^\gamma \psi \nn \\
&& - \epsilon_{\alpha \beta
\gamma}{\lambda^2 \over
  f^2}\oln{\psi}\varphi^\alpha(\vslash{\pd}\varphi^\beta)\tau^\gamma \psi
+ {2 m \lambda^2 \over f^2}\ora{\varphi}^2 \oln{\psi} \psi + \cdots,
\eqlab{lagrn1_new}
\eea
while the lagrangians $\mathcal{L}_{\pi \pi}^{(2)}$ 
and $\mathcal{L}_{\pi N}^{(2)}$ 
in Eqs.~(\ref{eq:lagr_pion_expl}, \ref{eq:lagr_nucl2_expl}) do not
change.
The transformed lagrangian, given by the sum of 
Eqs.~(\ref{eq:lagr_pion_expl}, \ref{eq:lagrn1_new}, \ref{eq:lagr_nucl2_expl}), 
defines a new representation of the theory; we will put tildes over
the quantities calculated from the transformed lagrangian.   

The tree-level isospin-symmetric amplitude in the new representation is the
sum of a new Born term, $\widetilde{D}^+_{Born} = \widetilde{D}^+_s +
\widetilde{D}^+_u$, and a new contact term, $\widetilde{D}^+_c$:
\beq
\widetilde{D}^+ = \widetilde{D}^+_{Born} + \widetilde{D}^+_c =
\widetilde{D}^+_s + \widetilde{D}^+_u + \widetilde{D}^+_c \,.
\eqlab{d_new_gen}
\eeq
Using the transformed lagrangian, we obtain for the new
s-channel, u-channel and contact diagrams:
\bea
\widetilde{D}^+_s &=& \frac{m g_A^2}{2 f^2} 
\left( \frac{\nu_B}{\nu_B - \nu} - {\nu \over 2m} \right) -
\frac{\lambda^2 (2m+\nu) + \lambda g_A \nu}{f^2}\,,
\eqlab{d_new_s}
\\
\widetilde{D}^+_u &=& \frac{m g_A^2}{2 f^2} 
\left( \frac{\nu_B}{\nu_B + \nu} + {\nu \over 2m} \right) -
\frac{\lambda^2 (2m-\nu) - \lambda g_A \nu}{f^2}\,,
\eqlab{d_new_u}
\\
\widetilde{D}^+_c &=& -{4 c_1 M^2 \over f^2} + {c_2 \over m^2 f^2}
\bigg[ (q p)(q' p) + (q p')(q' p') \bigg] + {2 c_3 \over f^2}(q q') 
+ {4 m \lambda^2 \over f^2}\,.
\eqlab{d_new_c}
\eea 
Note that both 
$\widetilde{D}^+_{Born} = \widetilde{D}^+_s + \widetilde{D}^+_u$ 
and $\widetilde{D}^+_c$ depend on the arbitrary transformation
parameter $\lambda$. Only the sum
$\widetilde{D}^+  = \widetilde{D}^+_{Born} +\widetilde{D}^+_c$ 
is independent of $\lambda$. Thus the complete tree-level scattering amplitude 
is invariant under the field transformation \eqref{transf} and we have:
\beq
\widetilde{D}^+ = D^+ \,.
\eqlab{invar_ampl}
\eeq

Exactly as was done in the original representation 
(see Eqs.~(\ref{eq:pion_se_gen}, \ref{eq:disp_eq})), the 
effective mass $\widetilde{M}_{eff}$ in the new representation 
is obtained by solving the dispersion equation
\beq
\omega^2 - M^2 - \widetilde{\Pi}(\omega,{\bf k}={\bf 0},\rho) = 0\,,
\eqlab{disp_eq_new}
\eeq
where the pion self-energy is proportional to the amplitude $\widetilde{D}^+$
evaluated at the forward scattering, nucleon rest frame kinematics:
\beq
\widetilde{\Pi}(\omega,{\bf k},\rho) = -\rho \,
\widetilde{D}^+(\nu=\omega,\nu_B=({\bf k}^2-\omega^2)/(2m))\,. 
\eqlab{pion_se_gen_new}
\eeq
Using Eqs.~(\ref{eq:d_new_gen}--\ref{eq:d_new_c}), we obtain
in the new representation:                                                            
\beq
\widetilde{M}_{eff}^2 = M^2 + \widetilde{M}_{eff, Born}^2 + 
\widetilde{M}_{eff, c}^2 + \mathcal{O}(Q^3,\rho^2)\,,
\eqlab{meff_new}
\eeq
where the Born and the contact term contributions equal
\beq
\widetilde{M}^2_{eff, Born} = \frac{\rho}{f^2} 
\left( \frac{g_A^2 M^2}{4m} + 4 m \lambda^2 \right)
\eqlab{meff_born_new}
\eeq
and
\beq
\widetilde{M}_{eff, c}^2=-{\rho \over f^2} \left[ \sigma + 2(c_2+c_3)M^2 
+ 4 m \lambda^2 \right]\,,
\eqlab{meff_ct_new}
\eeq
respectively. 
The $\lambda$-dependent terms present in $\widetilde{M}^2_{eff, Born}$ and
$\widetilde{M}^2_{eff, c}$ mutually cancel when the sum in \eqref{meff_new}
is taken, and comparison with \eqref{meff} shows that
the equivalence theorem is fulfilled for the effective mass:
\beq
\widetilde{M}_{eff} = M_{eff} \,,
\eqlab{invar_mass}
\eeq
which is a direct consequence of \eqref{invar_ampl}.
The crucial observation is that 
this representation invariance of the in-medium pion mass 
holds only provided {\em both} the Born and the contact 
contributions are 
taken into account. 
 
Without considering the nucleon field transformation,
one might falsely conclude from \eqref{meff} that the effective 
mass could be calculated based on the contact term alone.                
For example, from the outset one might want to use
$c_2' = c_2 - g_A^2/(8 m)$
instead of $c_2$ in the lagrangian $\mathcal{L}_{\pi N}^{(2)}$ of 
\eqref{lagr_nucl2_expl} in order to
forgo the Born term throughout the calculation. 
In doing so, one would have to take into
account the following two points.
First, the consistency of calculation would require one to drop the Born term in 
{\em all} considered representations.
Second, since the equivalence theorem dictates~\cite{Chi61,Lam73} that transitions between 
representations
are effected solely by changes of field variables while the
coupling constants must stay unchanged,
the constant $c_2$ could not be redefined differently in different representations.
In the original representation, a calculation based on the 
contact term with the constant $c'_2$ would lead to an effective mass of the form  
\beq
M^2 \left[ 1 - \frac{\rho}{f^2 M^2 } \left( \sigma +
2(c_2'+c_3)M^2 \right) \right] + \mathcal{O}(Q^3,\rho^2)\,,
\eqlab{meff_nob_r1}
\eeq
which numerically gives the same value as \eqref{meff}. However,
an effective mass calculated solely from the                        
corresponding contact term in the new representation would then equal
\beq
M^2 \left[ 1 - \frac{\rho}{f^2 M^2 } \left( \sigma +
2(c_2'+c_3)M^2 \right) \right] - 
\rho \frac{4m\lambda^2}{f^2} + \mathcal{O}(Q^3,\rho^2)\,.
\eqlab{meff_nob_r2}
\eeq
This expression contains an arbitrary $\lambda$-dependent term 
$\,-\rho \,4m \lambda^2/f^2$ and hence cannot in general 
be equal to \eqref{meff_nob_r1}.
Thus any attempt to mimic the Born contribution
by modifying the contact term will lead to                                        
an effective mass which will depend on an unphysical 
parameter, which is unacceptable since $M_{eff}$ is an
observable quantity that should not depend on the choice of representation.

\section{Concluding remarks}\seclab{concl}

In an approach based on the relativistic chiral lagrangian,
we have shown that it is impossible to have a 
representation-invariant in-medium pion mass
by considering the pion-nucleon contact interactions alone.  
We have also demonstrated that the representation invariance cannot be obtained   
by subsuming the nucleon-exchange effects (as represented by the Born term)
into the contact term.
Only by explicitly including the Born contribution, together with the
contact term, does the effective mass become invariant 
under redefinitions of the interpolating nucleon field.

As a physical observable, the effective mass should 
be independent of the choice of all interpolating fields of the lagrangian. 
The relativistic chiral lagrangian
employed in the present work at tree level provides a suitable model for 
studying effects of nucleon field transformations. 
It would be interesting to obtain analogous results within the framework of 
non-relativistic $\chi$PT which up to now has been
applied only in connection with meson field transformations~\cite{Yab94,Tho95,Lee95,Par02}. 
In contrast to the relativistic approach with the
pion-nucleon lagrangians Eqs.~(\ref{eq:lagr_nucl1_expl}, \ref{eq:lagr_nucl2_expl}) 
in the original representation, or
with their counterparts Eqs.~(\ref{eq:lagrn1_new}, \ref{eq:lagr_nucl2_expl}) 
in the new representation, the non-relativistic
heavy-baryon formalism using the standard dimensional regularization
would allow for a one-to-one correspondence between the loop and small 
momentum expansions~\cite{Gas88,Jen91,Ber93}.
Another line of development would be to consider consequences of simultaneous changes of
meson and nucleon field variables.
 
To make our argument as transparent as possible,
we adopted in the present paper the mean-field approximation limited 
to leading order in the nuclear density and considered s-wave interactions only. 
Nevertheless, our main conclusion -- that in calculating in-medium meson properties 
it is necessary to include the meson-nucleon Born contribution explicitly in
addition to the contact interaction -- follows from quite general
field-theoretical considerations and should therefore hold in more realistic models as well.

\begin{acknowledgments}

We thank Youngman Kim for useful discussions.
This work is supported in part by the US National Science Foundation,
Grant Nos.~PHY-9900756 and PHY-0140214. 

\end{acknowledgments}


\end{document}